\documentclass[aps,prl,twocolumn,]{revtex4}

\usepackage{graphicx}
\usepackage{amsmath}
\usepackage{bm}

\newcommand{\be}{\begin{equation}}
\newcommand{\ee}{\end{equation}}
\newcommand{\bwt}{\begin{widetext}}
\newcommand{\ewt}{\end{widetext}}
\newcommand{\bea}{\begin{eqnarray}}
\newcommand{\eea}{\end{eqnarray}}

\begin{document}
\title{Decoherence due to elastic Rayleigh scattering}

\author{H.~Uys,$^{1,2}$\protect\footnote{Electronic address:
\texttt{huys@csir.co.za}} M.J.~Biercuk,$^{1,3}$ A.P.~VanDevender,$^1$
C.~Ospelkaus,$^1$ D.~Meiser,$^4$ R.~Ozeri,$^5$ and
J.J.~Bollinger$^{1}$\protect\footnote{Electronic address:
\texttt{john.bollinger@nist.gov}}}
\affiliation{\mbox{$^1$National Institute of Standards and Technology, Boulder, CO 80305, USA}\\
\mbox{$^2$Council for Scientific and Industrial Research, Pretoria, South Africa}\\
\mbox{$^3$School of Physics, Univ. Sydney, Australia}\\
\mbox{$^4$JILA and Department of Physics, University of Colorado, Boulder, CO 80309-0440, USA}\\
\mbox{$^5$Dept. of Physics of Complex Systems, Weizmann Institute of Science,
Rehovot, Israel}}

\begin{abstract}
We present theoretical and experimental studies of the decoherence of hyperfine
ground-state superpositions due to elastic Rayleigh scattering of light
off-resonant with higher lying excited states.  We demonstrate that under
appropriate conditions, elastic Rayleigh scattering can be the dominant source
of decoherence, contrary to previous discussions in the literature.  We show
that the elastic-scattering decoherence rate of a two-level system is given by
the square of the difference between the elastic-scattering \textit{amplitudes}
for the two levels, and that for certain detunings of the light, the amplitudes
can interfere constructively even when the elastic scattering \textit{rates}
from the two levels are equal. We confirm this prediction through calculations
and measurements of the total decoherence rate for a superposition of the
valence electron spin levels in the ground state of $^9$Be$^+$ in a 4.5 T
magnetic field.
\end{abstract}

\maketitle


Off-resonant light scattering (spontaneous emission) is an important source of
decoherence in many coherent-control experiments with atoms and molecules.
Examples include the use of optical-dipole forces for gates in quantum
computing \cite{blattwineland08}, the generation of spin squeezed states
through laser-mediated interactions \cite{appj09,leri10,chas06,tepi08,gerj06},
and the trapping and manipulation of neutral atoms in optical lattices
\cite{kath03,misb10}. These experiments frequently involve superpositions of
two-level atomic systems (qubits) and use laser beams off-resonant with higher
lying excited states to control and measure the atomic states.

In general, decoherence of an atomic superposition state due to off-resonant
light scattering occurs if the scattered photon carries information about the
qubit state.  During Raman scattering the initial and final qubit states
differ.  The state of the scattered photon is entangled with the atomic state,
providing ``welcher-weg'' (which-way) information and leading to decoherence
\cite{moed04,ozer05}.  By contrast the role of elastic Rayleigh scattering for
decoherence is not as clear.  Two very different regimes have been discussed
and are supported by experiment. On the one hand it has been found that in some
experiments Rayleigh scattering gives rise to negligible decoherence provided
that the elastic scattering rates from both qubit levels are approximately
equal~\cite{ozer05}. On the other hand, decoherence due to Rayleigh scattering
of photons on a cycling transition is used for strong projective state
measurement~\cite{myea08}.

In this letter we develop a microscopic theory for the decoherence of a qubit
due to elastic Rayleigh scattering that gives a unified treatment of these
different regimes. Our key finding is that the decoherence induced by Rayleigh
scattering is proportional to the square of the difference of the probability
\emph{amplitudes} for elastic scattering from the two levels
[Eq.~(\ref{gammadduunew})].  When the two amplitudes are approximately equal
the resulting decoherence rate can be small (first case above) and when one
amplitude dominates the other, Rayleigh decoherence can be large (second case
above).  However, even when the elastic scatter rates are approximately equal,
the scattering amplitudes can have opposite sign -- a situation applicable to
many coherent control experiments \cite{appj09,leri10,chas06,tepi08,gerj06} --
and Rayleigh decoherence can dominate Raman decoherence.

\begin{figure}
\includegraphics[width=6.5 cm,clip]{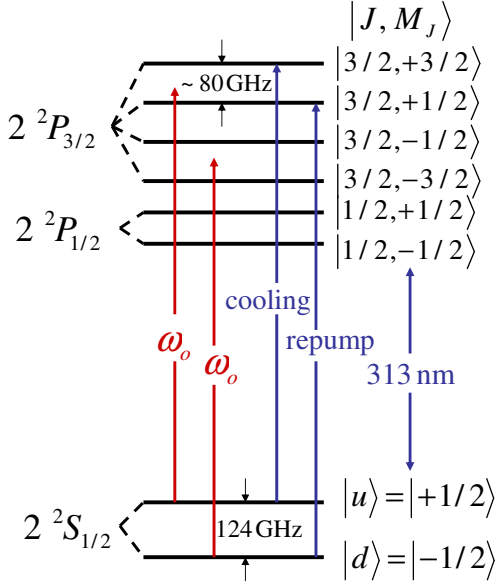}
\caption{Relevant energy level structure of $^9$Be$^+$ at 4.5 T. We only show
the $m_{I}=+\frac{3}{2}$ levels which are prepared experimentally though
optical pumping. The off-resonant laser is shown with a frequency $\omega_{o}$
appropriate for the experimental results presented in
Fig.~\ref{results}.}\label{Belevels}
\end{figure}

In the rest of this paper we derive the microscopic model that quantitatively
describes the decoherence due to Rayleigh scattering and verify our predictions
against an experiment with ${}^9\text{Be}^+$ in a 4.5 T magnetic field.  For
concreteness we discuss the problem in terms of the relevant energy level
structure of $^9$Be$^+$ shown in Fig.~\ref{Belevels}.  The two ground state
sub-levels $|u\rangle = |+\frac{1}{2}\rangle$ and $|d\rangle = |-\frac{1}{2}
\rangle$ corresponding to the valence electron spin states parallel and
anti-parallel to an applied magnetic field are split by $\Omega_z$ =
2$\pi\times$124.1 GHz.  Also shown are the first excited P state levels
$|J,M_{J}\rangle$, $J=\frac{3}{2},\frac{1}{2}$. We consider a non-resonant
light field $(\sum_{\lambda=-1}^{1} b_{\lambda}\hat\epsilon_{\lambda})
\mathcal{E}_{0} \cos(\mathbf{k}_0\cdot\mathbf{x}-\omega_0t)$, where
$b_{\lambda}$ is the amplitude of the corresponding polarization component
$\hat\epsilon_{\lambda}$ of the light, and derive a master equation for the
time evolution of the $|\pm\frac{1}{2} \rangle$ ground state sub-levels in the
presence of the off-resonant light.

An effective Hamiltonian coupling of the two qubit levels via spontaneous
scattering of the driving field into the vacuum modes can be derived by
adiabatic elimination of the excited states shown in Fig.~\ref{Belevels} and is
given by \be H_{sp}=H_{dd}+H_{uu}+H_{du}+H_{ud}.\label{Hsp}\ee Here the
different terms describe processes which either do, or do not, flip the atomic
spin upon scattering of a photon, as indicated by the subscripts. The Raman
terms that lead to spin-flips are (in the Schr\"odinger picture), \small \bea
H_{du}\!\!&=&\!\!\hbar\sum_{J\mathbf{k}\lambda}
\left(h^{du}_{J\mathbf{k}\lambda} \hat\sigma^-\hat
a_{\mathbf{k}\lambda}E^{-}(\Delta\mathbf{k})+h^{du*}_{J\mathbf{k}\lambda}\hat
\sigma^+\hat a^\dagger_{\mathbf{k}\lambda}E^{+}(\Delta\mathbf{k})\right)\nonumber\\
H_{ud}\!\!&=&\!\!\hbar\sum_{J\mathbf{k}\lambda}
\left(h^{ud}_{J\mathbf{k}\lambda}\hat \sigma^-\hat
a^\dagger_{\mathbf{k}\lambda}E^{+}(\Delta\mathbf{k})+h^{ud*}_{J\mathbf{k}\lambda}\hat
\sigma^+\hat
a_{\mathbf{k}\lambda}E^{-}(\Delta\mathbf{k})\right),\nonumber\label{Huddu} \eea
\normalsize where $\hat a^\dagger_{\mathbf{k}\lambda}$ ($\hat
a_{\mathbf{k}\lambda}$) is the raising (lowering) operator for the
$\mathbf{k},\lambda$ mode of the electromagnetic field, $\hat \sigma^{+} =
(\hat \sigma_{x} + i\hat \sigma_{y})/2$ and $\hat \sigma^{-} = (\hat \sigma_{x}
- i\hat \sigma_{y})/2$ where $\sigma_{x}$, $\sigma_{y}$, and $\sigma_{z}$ are
the Pauli matrices for the two-level ground state, and
\small$E^{+}(\Delta\mathbf{k})\!=E^{-*}(\Delta\mathbf{k})\!=(\mathcal{E}_0/2)e^{i(\mathbf{k}_0-\mathbf{k}
)\cdot\mathbf{x}-i\omega_0t}$. \normalsize The coupling constants are \small
\be h^{ij}_{J\mathbf{k}\lambda}\!\!\!=\!\!-\frac{b_\lambda
g_\mathbf{k}}{\hbar^2\delta_{iJ\lambda}}\langle
i|\mathbf{d}\cdot\hat\epsilon_\lambda^*|J, \lambda+i\rangle\langle
J,\lambda+i|\mathbf{d}\cdot\hat\epsilon_{\lambda+(i-j)}|j\rangle,\nonumber
\ee \normalsize where $i,j \in \{-\frac{1}{2}, \frac{1}{2}\}$,
$\delta_{iJ\lambda}=\Omega_{J,\lambda+i}-\Omega_{i}-\omega_0$ is the detuning
of $\omega_0$ from the $|i\rangle\rightarrow|J, M_{J}=\lambda+i\rangle$
transition, at energies $\hbar\Omega_{i}$ and $\hbar\Omega_{J,\lambda+i}$
respectively, and \small
$g_\mathbf{k}\!=\!\sqrt{\hbar\omega_\mathbf{k}/(2\epsilon_0V)}$ \normalsize
with $V$ the quantization volume. Similarly the Rayleigh terms are \small \bea
H_{dd}\!\!&=&\!\!\hbar\sum_{J\mathbf{k}\lambda}
\left(h^{dd}_{J\mathbf{k}\lambda}\hat
a_{\mathbf{k}\lambda}E^{-}(\Delta\mathbf{k})+h^{dd*}_{J\mathbf{k}\lambda}\hat
a^\dagger_{\mathbf{k}\lambda}E^{+}(\Delta\mathbf{k})\right)\hat\sigma_1\nonumber\\
H_{uu}\!\!&=&\!\!\hbar\sum_{J\mathbf{k}\lambda}
\left(h^{uu}_{J\mathbf{k}\lambda}\hat
a_{\mathbf{k}\lambda}E^{-}(\Delta\mathbf{k})+h^{uu*}_{J\mathbf{k}\lambda}\hat
a^\dagger_{\mathbf{k}\lambda}E^{+}(\Delta\mathbf{k})\right)\hat\sigma_2,\nonumber
\eea \normalsize where we've defined the operators
$\hat\sigma_1\!=\!\frac{1}{2}(I-\sigma_z)\!=\!|d\rangle\langle
d|,\,\,\hat\sigma_2\! =\!\frac{1}{2}(I+\sigma_z)\!=\!|u\rangle\langle u|$. With
these definitions the Rayleigh terms can be combined $H_{el}=H_{dd}+H_{uu}$
where \bea H_{el}\!\!&=&\!\!\frac{\hbar}{2}\sum_{J\mathbf{k}\lambda}
\left(h^{dd}_{J\mathbf{k}\lambda}+h^{uu}_{J\mathbf{k}\lambda}\right)\hat
a_{\mathbf{k}\lambda}E^{-}(\Delta\mathbf{k}
)\nonumber\\
-\frac{\hbar}{2}\hat\sigma_z\!\!\!\!&\sum\limits_ {J\mathbf{k}\lambda}&\!\!\!\!
\left(h^{dd}_{J\mathbf{k}\lambda}-h^{uu}_{J\mathbf{k}\lambda}\right)\hat
a_{\mathbf{k}\lambda}E^{-}(\Delta\mathbf{k} ) + H.C..\label{Hdduu} \eea The
first term in Eq.~(\ref{Hdduu}) is proportional to the identity for the qubit
and consequently does not lead to decoherence.  The second term proportional to
$\hat\sigma_z$ leads to pure dephasing which, as we will see, can sometimes be
the dominant decohering process.

The time evolution of the reduced density matrix, $\tilde\rho_S(t^\prime)$, of
the qubit is well described by the master equation \cite{gardinerzoller} (in
the interaction picture) \be \frac{\partial \tilde\rho_S(t)}{\partial t}=
-\frac{1}{\hbar^2}\int^t_0dt^\prime\textnormal{Tr}_\mathbf{k}\left\{\left[H_{sp}(t),\left[H_{sp}(t^\prime),
\rho_S(t^\prime)\rho_\mathbf{k}\right]\right] \right\} , \label{mastereq} \ee
in which $\rho_S(t^\prime)$ is the density matrix of the qubit before tracing
over all vacuum modes represented by density matrix $\rho_\mathbf{k}$ as
indicated by Tr$_\mathbf{k}$.

In evaluating Eq.~(\ref{mastereq}) we find Raman scattering terms given in the
Lindblad form by \footnote{When multiplying out the double commutator in
Eq.~(\ref{mastereq}), products between different Hamiltonians ($H_{el}$,
$H_{ud}$, or $H_{du}$) are neglected because they are rapidly oscillating of
order $\Omega_{z}$ (secular approximation).} \small \bea
\mathcal{L}_{du}\tilde\rho_S(t)&=&-\frac{\Gamma_{du}}{2}\left(
\hat\sigma^{-}\hat\sigma^{+}\tilde\rho_S(t)-2\hat\sigma^{+}\tilde\rho_S(t)\hat\sigma^{-}
+\tilde\rho_S(t)\hat\sigma^{-}\hat\sigma^{+}\right)\nonumber\\
\mathcal{L}_{ud}\tilde\rho_S(t)&=&-\frac{\Gamma_{ud}}{2}\left(
\hat\sigma^{+}\hat\sigma^{-}\tilde\rho_S(t)-2\hat\sigma^{-}\tilde\rho_S(t)\hat\sigma^{+}
+\tilde\rho_S(t)\hat\sigma^{+}\hat\sigma^{-}\right).\nonumber\label{Lrhoud}
\eea \normalsize $\Gamma_{ij}$ is the rate for an ion initially in state
$|i\rangle$ to scatter a photon and end up in state $|j\rangle$ and is given by
the Kramers-Heisenberg formula
\begin{equation}
\Gamma_{ij}={\Omega_R}^2\gamma\sum_{\lambda}\left(\sum_JA^{i\rightarrow
j}_{J,\lambda}\right)^2.
\end{equation}
Here $\Omega_R=\mu\mathcal{E}_0/(2\hbar)$, $\gamma$ is the spontaneous decay
rate of the excited states, and the amplitudes $A^{i\rightarrow j}_{J,\lambda}$
for scattering through intermediate level $|J,\lambda+i\rangle$ are defined as
\begin{equation}
A^{i\rightarrow j}_{J,\lambda}=\frac{b_{\lambda}\langle
j|\mathbf{d}\cdot\hat\epsilon_{\lambda+(i-j)}^{*}|J,\lambda+i\rangle\langle
J,\lambda+i|\mathbf{d}\cdot\hat\epsilon_{\lambda}|i\rangle}{\delta_{iJ\lambda}\,\mu^2}
\;\label{matrix_el}
\end{equation}
where
$\mu=\langle\frac{3}{2},\frac{3}{2}|\mathbf{d}\cdot\hat\epsilon_{1}|\frac{1}{2}\rangle$
is the largest dipole matrix element. We define the total Raman scattering rate
$\Gamma_{Ram}=\Gamma_{du}+\Gamma_{ud}$.

The contributions to Eq.~(\ref{mastereq}) from Rayleigh scattering are \be
\mathcal{L}_{dd,uu}\tilde\rho_S(t)=-\frac{\Gamma_{el}}{4}\left[
\tilde\rho_S(t)-\hat\sigma_z\tilde\rho_S(t)\hat\sigma_z\right]\label{Lrhodduunew}
\ee where \be
\Gamma_{el}=\Omega_R^2\gamma\sum_\lambda\left(\sum_{J}A^{d\rightarrow
d}_{J,\lambda}-\sum_{J^\prime}A^{u\rightarrow
u}_{J^\prime,\lambda}\right)^2.\label{gammadduunew} \ee Finally the full time
evolution of the reduced density matrix is $\frac{\partial
\tilde\rho_S(t)}{\partial t} = \left( \mathcal{L}_{du}+ \mathcal{L}_{ud}+
 \mathcal{L}_{dd,uu}\right)\tilde\rho_S(t)$.
With
$\tilde\rho_S(t) \equiv\left( \begin{array}{cc} \rho_{uu} & \rho_{ud} \\
\rho_{du} & \rho_{dd}\end{array} \right)$ this can be written

\be \frac{\partial \tilde\rho_S(t)}{\partial t} = \left(
\begin{array}{cc}
-\Gamma_{ud}\rho_{uu}+\Gamma_{du}\rho_{dd} & -\frac{1}{2}(\Gamma_{Ram}+\Gamma_{el})\rho_{ud} \\
 -\frac{1}{2}(\Gamma_{Ram}+\Gamma_{el})\rho_{du}&-\Gamma_{du}\rho_{dd}+\Gamma_{ud}\rho_{uu}
\end{array} \right). \label{matrixeq}\ee

Eq.~(\ref{gammadduunew}) is the main theoretical result of this letter. General
master equation treatments of decoherence due to elastic scattering have
appeared in the literature \cite{cohc92}.  However, to our knowledge this is
the first calculation producing a simple formula which demonstrates the central
role of elastic scattering amplitudes in spontaneous-emission-induced
decoherence of hyperfine-state superpositions.

Since the detunings in the two amplitudes in Eq.~(\ref{gammadduunew}) can have
opposite sign, the different contibutions can add constructively, leading to a
large Rayleigh scattering decoherence. Physically the terms proportional to
$\hat \sigma_{z}$ in Eq.~(\ref{Hdduu}) suggest that this decoherence is due to
dephasing. A photon elastically scattered into mode $\mathbf{k},\lambda$
produces a shift in the phase of a superposition of the $|u\rangle, |d\rangle$
levels proportional to the difference in the amplitudes for scattering into
that mode. Since photon scattering occurs at random times and into random
directions, random phase shifts are produced (due to the
\small$E^{+}(\Delta\mathbf{k})$ \normalsize coefficient), and consequently
phase information of the superposition state is lost.

\begin{figure}
\includegraphics[width=8.5 cm,clip]{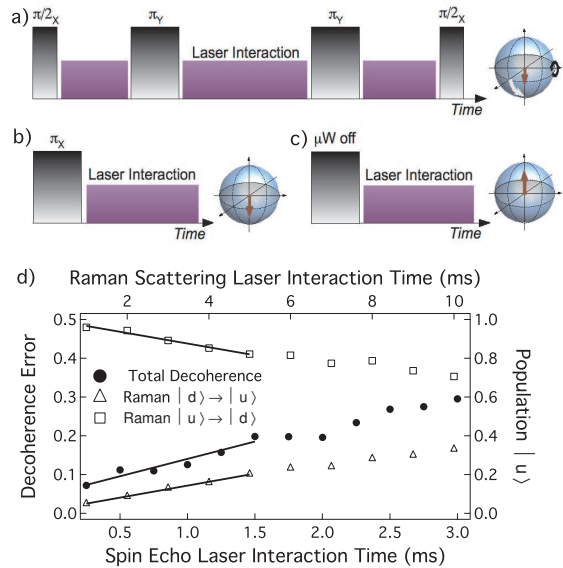}
\caption{(a-c) Schematic pulse sequences used to measure (a) total decoherence
and (b), (c) Raman scattering rates.  Raman scatter rates $\Gamma_{du}$ and
$\Gamma_{ud}$ (used to calibrate laser intensity, see Fig.~\ref{results})
obtained by initializing the ions in $|d\rangle$ or $|u\rangle$ respectively
and measuring the $|u\rangle$ state population vs off-resonant laser
interaction time. (d) Example measurements of the total decoherence and Raman
scatter rates.  Total decoherence (solid markers), left axis; Raman scatter
rates (open markers), right axis.}\label{techniques}
\end{figure}

We realize elastic scattering amplitudes with opposite sign in a Penning ion
trap operating in the strong-field Zeeman regime. These traps have found
increasing importance in quantum information studies
\cite{crid10,biem09a,uysh09}, underscoring the need for quantitative analysis
of decoherence processes in these systems. Details of our basic Penning trap
set-up can be found in \cite{biem09b}.  We use two 313 nm laser beams (see
Fig.~\ref{Belevels}) to Doppler laser cool (T$\sim$0.5 mK) and optically pump a
planar crystal (diameter $\sim$300 $\mu$m) of $\sim$100 $^9$Be$^{+}$ ions  into
the $|u\rangle=|+1/2\rangle$ state. A phase-locked Gunn diode oscillator near
124 GHz provides microwave radiation used to perform high-fidelity ($>$99 \%)
qubit rotations.  At the end of an experiment, the population in the
$|u\rangle$ state is measured by detecting the resonantly scattered ion
fluorescence induced by the Doppler cooling laser (tuned to the
$|u\rangle\rightarrow|3/2,3/2\rangle$ cycling transition), and normalizing this
fluorescence level to that obtained with all the ions in $|u\rangle$.

We introduce an off-resonant, linearly polarized laser beam and measure the
total decoherence rate $\frac{1}{2}(\Gamma_{Ram}+\Gamma_{el})$ from a decrease
in length of the composite Bloch vector of the ions using a two-$\pi$-pulse
spin echo sequence like that shown in Fig.~\ref{techniques}(a).  We adjust the
relative phases of the $\pi$/2 and $\pi$ pulses so that in the absence of any
coherence loss all ions are rotated to $|d\rangle$ at the end of the spin-echo
sequence.  The measured population in $|u\rangle$ therefore provides a measure
of the total decoherence. From Eq.~(\ref{matrixeq}) we calculate that at the
end of the spin echo sequence
$\rho_{uu}=\frac{1}{2}(1-e^{-(\Gamma_{Ram}+\Gamma_{el})\tau/2})$, where $\tau$
is the total laser interaction time of the spin echo sequence.
Figure~\ref{techniques}(d) shows measurements of the $|u\rangle$ state
population as a function of the spin echo laser interaction time. We determine
$\frac{1}{2}(\Gamma_{Ram}+\Gamma_{el})$ from twice the fitted slope at short
spin-echo sequence times.

Other sources of decoherence can be neglected. Without off-resonant light we
measure a $\sim1~\%$ loss in coherence due to magnetic field fluctuations for
up to 6 ms spin echo times, much longer than the spin echo sequence times used
to determine decoherence due to off-resonant light scattering. Further, the
ac-stark shift from the off-resonant laser is nulled to better than 1 kHz by
rotating the polarization. Doing so minimizes decoherence due to laser-power
fluctuations during the spin-echo probe sequence. Electric field homogeneity
due to the laser beam waist is greater than 95$~\%$ over the ion planar array.

\begin{figure}
\includegraphics[width=8.5 cm,clip]{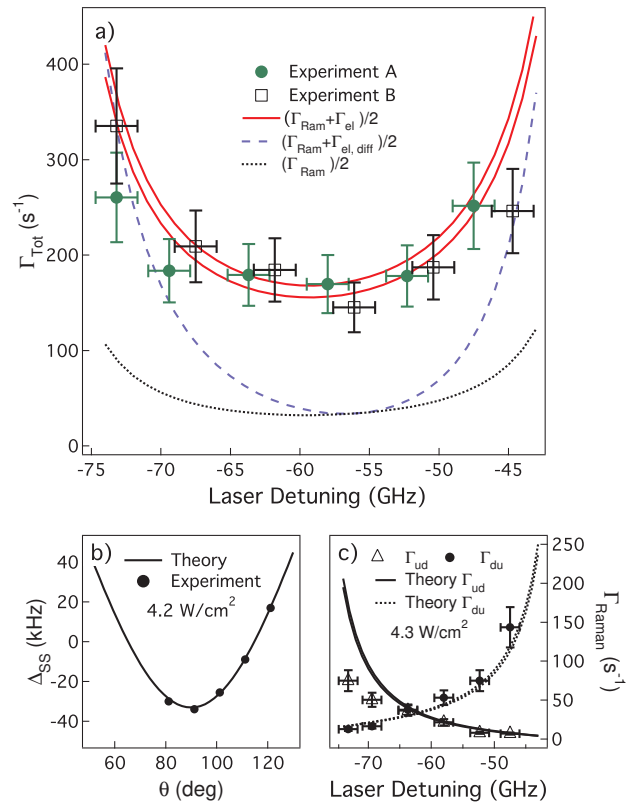}
\caption{(Color online) (a) Comparison of the total decoherence rate with
different theoretical models as a function of light detuning from the
$|u\rangle\to |3/2,3/2\rangle$ cycling transition for two independent sets of
measurements.  The light intensity was calibrated through fits to (b) light
shift measurements of the qubit transition vs the polarization angle
(Experiment A) and (c) Raman scatter rates (Experiment B).  The full theory
developed in this paper agrees with the experimental data for all detunings
while a theory accounting only for decoherence due to Raman transitions
significantly underestimates the decoherence rate and an estimate based on
scattering rate differences ($\Gamma_{el,diff}$) agrees with experiment only
near the $|u\rangle\rightarrow|3/2,1/2\rangle$ and
$|d\rangle\rightarrow|3/2,-1/2\rangle$ resonances at -79.4 and -37.7 GHz,
respectively. The difference in the two full theory curves is due to
uncertainty in the laser polarization and absolute laser intensity calibration.
}
\label{results}
\end{figure}

Figure~\ref{results}(a) presents two sets of measurements of the total
decoherence rate as a function of the frequency of the off-resonant light, and
compares these measurements with different theoretical predictions.  For these
measurements the intermediate state detunings and resulting $d\rightarrow d$
and $u\rightarrow u$ scattering amplitudes in Eq. (\ref{gammadduunew}) have
opposite sign, producing a large elastic decoherence rate.  This condition is
in part achieved because the light detunings to the intermediate $P_{3/2}$
levels are comparable to the large (124 GHz) qubit splitting.

The two measurement sets use different ion crystals and different techniques
for calibrating the off-resonant light intensity (see
Figs~\ref{results}(b),(c)). Good agreement is obtained with the full theory
presented here which manifests a Rayleigh decoherence rate determined by the
square of the difference between the $|u\rangle$ and $|d\rangle$ elastic
scattering \textit{amplitudes} (Eq.~(\ref{gammadduunew})). This contrasts with
the idea conveyed in previous literature that decoherence due to elastic
scattering is produced by a difference in elastic scattering \textit{rates}.
For example Ref. \cite{ozer07} (Sec. III.B) discusses the which-way information
acquired due to a difference in Rayleigh scattering rates and estimates a
decoherence rate $\frac{1}{2}\Gamma_{el,diff}=
\frac{(\Gamma_{uu}-\Gamma_{dd})^2}{(\Gamma_{uu}+\Gamma_{dd})}$ due to these
rate differences.
Figure~\ref{results}(a) shows theory based on $\Gamma_{el,diff}$. As the laser
detuning approaches resonant transitions at $-79.4$ GHz and $-37.7$ GHz,
reasonable agreement is obtained with this theory. However, at a detuning of
$\sim -56$ GHz, $\Gamma_{uu}\approx\Gamma_{dd}$ and we measure a decoherence
rate that is five times larger than that predicted by $\frac{1}{2}
(\Gamma_{Ram} + \Gamma_{el,diff})$.

These measurements can be contrasted with previous work \cite{ozer05} using
trapped $^{9}$Be$^{+}$ ions at low magnetic field in which
Rayleigh-scattering-induced decoherence was shown to be negligible. The results
of that experiment benefited from some fortuitous conditions: the detuning
$\Delta$ of the light from any atomic excited state was large compared to the
qubit transition frequency, and the qubit states were clock states (states
whose energy difference does not depend to first order on changes in the
magnetic field). These conditions imply a nearly complete cancellation of the
elastic scattering amplitudes in Eq. (\ref{gammadduunew}) resulting in a weak
Rayleigh decoherence with a dependence on detuning $\propto 1/\Delta^{4}$.  For
general two-level systems the cancellation will not be as complete, resulting
in a Rayleigh decoherence rate $\propto 1/\Delta^{2}$.  The prescription
discussed in this manuscript now enables an accurate calculation of Rayleigh
decoherence for these low-field trapped ion as well as other coherent control
experiments.


We thank W. M. Itano, J. P. Britton, D. Hanneke, and M. J. Holland for useful
suggestions. MJB acknowledges fellowship support from Georgia Tech and IARPA.
DM is supported by NSF. This work was supported by the DARPA OLE program and by
IARPA. This manuscript is the contribution of NIST and is not subject to U.S.
copyright.


\end{document}